# Silicon Photonic Direct-Detection Phase Retrieval Receiver

Brian Stern, Haoshuo Chen, Kwangwoong Kim, Hanzi Huang, Jie Zhao, Mohamad Hossein Idjadi

Nokia Bell Labs, Murray Hill, NJ, USA

**Abstract** *We demonstrate a direct-detection phase retrieval receiver based on silicon photonics. The receiver implements strong dispersion and delay lines on a compact chip. We retrieve the full field of a 30-GBd QPSK signal without a carrier or local oscillator. ©2023 The Authors*

**Introduction**
The bandwidth requirements for optical interconnects in data centers, passive optical networks (PONs), metro and long-haul networks highlight the need for low-cost devices [1–4]. Higher bit rates in these applications suggest coherent optics as a possible solution. However, coherent receivers typically require lasers to act as local oscillators (LO), adding to the receiver cost and introducing a possible point of failure. Other approaches, such as the self-coherent Kramers-Kronig and Stokes-vector receivers, experience a reduced receiver sensitivity compared to conventional coherent detection since a large amount of power is used by the carrier which is co-transmitted with the signal [5–7].

An alternative full-field receiver architecture was proposed using direct detection [8–11], removing the need for an LO or carrier, or for precise wavelength alignment of the upstream signals. This receiver scheme, based on space-time diversity phase retrieval, was demonstrated using optical fibers and bulk modules. However, integrated systems would better leverage large-scale deployment and enable lower costs, and silicon photonics has emerged as a key platform for integrated coherent receivers [12,13]. There are also several recent demonstrations of integrated self-coherent receivers based on direct detection, though they still require transmitting a carrier [14–16]. Our recent work explored increasing dispersion on-chip for carrier-less phase retrieval applications [17], but an integrated receiver based on phase retrieval has not been shown.

Here we demonstrate a fully-integrated, carrier-less silicon photonic phase retrieval receiver. The chip is based on microring resonators, silicon waveguide delay lines, and integrated high-speed photodiodes. We show recovery of a 30-GBd quadrature phase shift keying (QPSK) signal using the direct-detection receiver and a phase retrieval algorithm, without an LO or transmitted carrier.

**Phase Retrieval with Space-Time Diversity**
Typical coherent receivers recover phase information from an incoming signal. In contrast, direct detection typically only recovers the signal's intensity but not phase. However, with a space-time diversity scheme at the receiver, phase retrieval may be accomplished with direct detection (see Fig. 1). The receiver consists of four intensity measurements: a set of varying time delays $\Delta t$ for symbol-to-symbol interference, and a dispersive element $D$ which relates phase to amplitude modulation. Based on these intensity measurements, the phase retrieval algorithm iteratively solves for the input phase, e.g., using a modified Gerchberg-Saxton algorithm [10,18].

**Integrated Dispersion**
We design a set of microring resonators to provide a large on-chip dispersion for one branch of the phase retrieval receiver. Ring resonators have been used in multistage filters for dispersion compensation [19]. Here we use a set of silicon microrings to ensure sufficient dispersion over a sufficient bandwidth for the receiver, with minimal group delay ripple. Figure 2(a) shows a diagram of the concept. Three rings, composed of single-mode silicon waveguides, are arranged in an all-pass configuration. They have identical circumferences corresponding to 30 µm radius, but their power coupling strengths κ vary, at 23%, 31%, and 39%, respectively. The ring with the lowest κ has the highest extinction ratio and largest peak group delay near resonance. The subsequent two rings contribute smaller delays. By detuning these two rings from the first (using integrated heaters), an approximately linear, decreasing group delay with respect to wavelength is achieved (Fig. 2(b)). The slope in this simulation corresponds to a dispersion of -94 ps/nm, which is sufficient for phase retrieval. The group delay ripple is only ~1 ps over the ~60 GHz bandwidth. Being overcoupled, the rings' low extinction ratio leads to a simulated transmission loss below 0.2 dB. The rings are designed to operate on the transverse-electric (TE) mode.

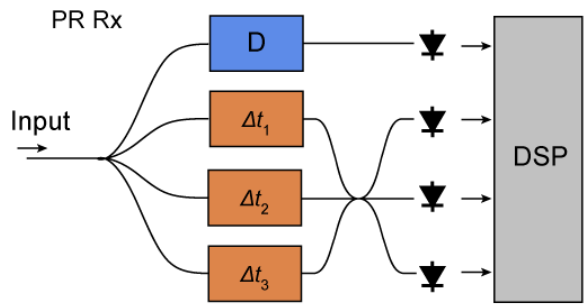

**Fig. 1:** Schematic of a phase retrieval (PR) receiver, including elements providing dispersion, time delays, photodiodes, and digital signal processing (DSP).

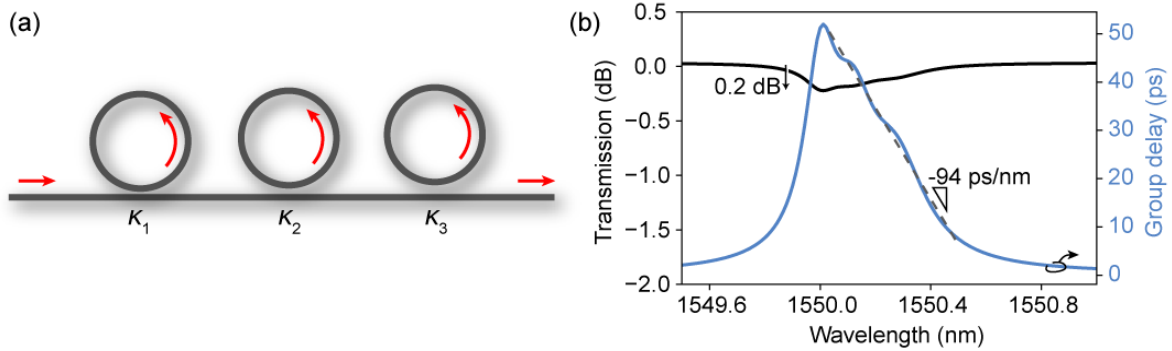

**Fig. 2:** On-chip dispersion using microring resonators. (a) Diagram of a set of microrings along a bus waveguide, (b) Simulation of the transmission and group delay of the set of rings.

**Integrated Delay Lines and Receiver Layout**
In order to implement the time-delayed branches of the phase retrieval receiver, we include long lengths of silicon waveguide delay lines. Figure 3 shows the delay lines in the context of the entire layout. The first branch has negligible delay length $L_1$. The next two branches have lengths $L_2$ = 1.3 cm and $L_3$ = 2.2 cm, corresponding to delays of ~120 ps and ~200 ps, respectively. By shaping the waveguides into spirals, such long lengths are accomplished in a small area on-chip. The delays produce symbol-to-symbol interference of 6 and 3.5 symbol periods at 30 GBd. To minimize propagation losses, we use extra-wide waveguides to reduce interaction with waveguide sidewall roughness. Additionally, the increased confinement increases the effective index of refraction and therefore the delay time. Because the waveguides are now multimode, we use Euler bends in the spirals to avoid exciting higher-order modes. We expect the total spiral delay branch losses to be approximately 1.5 dB and 2.5 dB, respectively. These may be improved using partially etched waveguides or other materials.

The received optical signal is split equally among the four branches with a 1x4 multimode interferometer (MMI) splitter. The three delay line branches are mixed using a 3x3 MMI. The simulated insertion losses of the MMIs are 0.14 dB and 0.13 dB, respectively, staying below 0.32 dB and 0.41 dB across the C band. The integrated germanium photodiodes have a bandwidth exceeding 40 GHz and cover 1500-1600 nm.

**Fabrication and Assembly**
The phase retrieval receiver is fabricated at a commercial foundry on 300 mm silicon-on-insulator (SOI) wafers with 220 nm device layer thickness. The process includes silicon waveguides, ion-implanted resistive heaters, germanium photodiodes, and metallization.

We assemble a fabricated chip measuring 1.7 mm × 1.4 mm to characterize the receiver performance (Fig. 4). We mount the chip on a thermo-electric cooler (TEC) and attach a lensed fiber coupled to the input waveguide (<2 dB loss). The chip's heater connections are wirebonded, while we use a high-speed probe for the photodiode measurements.

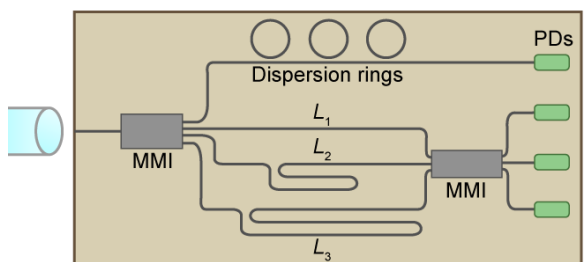

**Fig. 3:** Diagram of the integrated phase retrieval receiver.

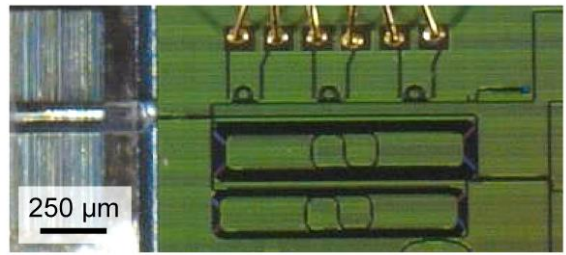

**Fig. 4:** Microscope photograph of the fabricated receiver chip.

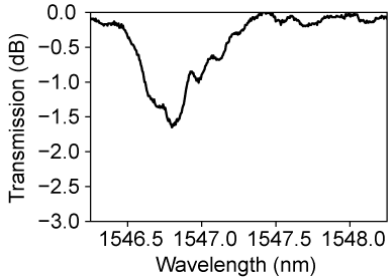

**Fig. 5:** OSA measurement of the rings' spectrum with the resonances properly detuned. The resonances repeat with a free spectral range of ~3 nm.

### Results

We first investigate the performance of the dispersion rings and determine proper tuning settings. The photonic integrated circuit (PIC) includes an optical tap to a grating coupler between the rings and their photodiode. Using this output monitor, we observe the transmission spectrum of this branch and identify resonances corresponding to each ring. The rings are tuned by three heaters with resistances of 1.3 kΩ, while the TEC is set to room temperature. The rings' resonances are detuned slightly to match the shape expected from Fig. 2b. The optical spectrum analyzer (OSA) trace of the transmission of the tuned resonances is shown in Fig. 5. The spectrum is largely as expected, but the extinction ratio is higher, which may lead to a larger dispersion. Still, the insertion loss across the ~50 GHz bandwidth is only about 1 dB.

We next demonstrate a phase retrieval receiver with a line rate of 60 Gb/s using the photonic integrated circuit. We use the experimental setup shown in Fig. 6(a). The output of an external-cavity laser is modulated by an in-phase and quadrature Mach-Zehnder modulator (IQ-MZM) driven by a two-channel digital-to-analog converter (DAC) using a Nyquist-shaped 30-GBd QPSK signal of length $2^{12}$. This is followed by an erbium-doped fiber amplifier (EDFA) and a 40-km equivalent dispersion compensating fiber. The polarization is aligned to TE at the chip, and the input power is 5.8 dBm. The heaters are aligned to the settings previously determined. The photodiodes are biased to -2 V and the electrical signals are captured at 80 GS/s using a multi-channel analog-to-digital converter (ADC). The signal is reconstructed off-line using a parallel, block-wise phase retrieval algorithm with a block length of 4,096 samples at 2 samples per symbol. Figure 6(b) shows that the reconstructed signal matches closely to the target signal. Next, the reconstructed QPSK signal's constellation is shown in Fig. 6(c). The calculated bit error rate (BER) is $8.6 \times 10^{-3}$, which is below

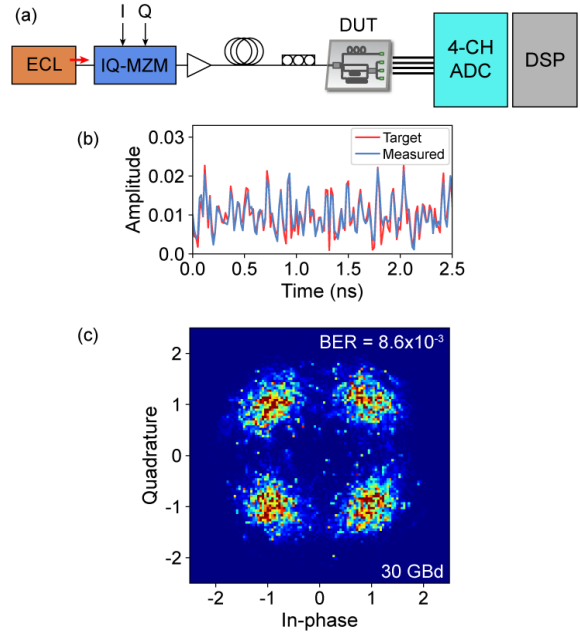

**Fig. 6:** (a) Diagram of the experimental setup for the phase retrieval receiver. ECL: external cavity laser, MZM: Mach-Zehnder modulator, (b) Measured reconstructed amplitude compared to the target signal, (c) Measured QPSK constellation at 30 GBd.

the 20% forward error correction (FEC) threshold.

### Conclusions

We have demonstrated a silicon photonic direct-detection phase retrieval receiver operating at 30 GBd. The receiver recovers QPSK-modulated signals without a carrier or LO by leveraging a space-time diversity scheme in conjunction with a phase retrieval algorithm. The realization of such a receiver on a compact photonic chip highlights the potential for low-cost direct-detection solutions for data centers, PON, metro and long-haul optical networks.

We expect improved performance is attainable by using lower-loss waveguide designs to enable longer delay lengths, and polarization diversity may also be implemented with typical silicon photonic building blocks.

### Acknowledgements
We wish to thank Andrea Blanco-Redondo, Nicolas K. Fontaine, Di Che and Xi Chen for fruitful discussions.